# Layer Dependent Interfacial Transport and Optoelectrical Properties of MoS$_2$ on Ultra-flat Metals

Hao Lee[1], S. Deshmukh[2], Jing Wen[3,4], V.Z. Costa[1], J. S. Schuder[5], M. Sanchez[5], A. S. Ichimura[5], Eric Pop[2,6,7], Bin Wang[3], and A. K. M. Newaz[1]

[1]Department of Physics and Astronomy, San Francisco State University, San Francisco, California 94132, USA
[2]Department of Electrical Engineering, Stanford University, Stanford, California 94305, USA
[3]School of Chemical, Biological and Materials Engineering, University of Oklahoma, Norman, Oklahoma 73019, United States
[4]Key Laboratory for Photonic and Electronic Bandgap Materials, Ministry of Education, School of Physics and Electronic Engineering, Harbin Normal University, Harbin 150025, P. R. China
[5]Department of Chemistry and Biochemistry, San Francisco State University, San Francisco, California 94132, USA
[6]Department of Materials Science and Engineering, Stanford University, Stanford, California 94305, USA
[7]Precourt Institute for Energy, Stanford University, Stanford, California 94305, USA

**Abstract**

Transition metal dichalcogenides (TMDs) are layered semiconducting van der Waal crystals and promising materials for a wide range of electronic and optoelectronic devices. Realizing practical electrical and optoelectronic device applications requires a connection between a metal junction and a TMD semiconductor. Hence, a complete understanding of electronic band alignments and the potential barrier heights governing the transport through a metal-TMD-metal junction is critical. But, there is a knowledge gap; it is not clear how the energy bands of a TMD align while in contact with a metal as a function of the number of layers. In pursuit of removing this knowledge gap, we have performed conductive atomic force microscopy (CAFM) of few layered (1-5) MoS$_2$ immobilized on ultra-flat conducting Au surfaces (root mean square (RMS) surface roughness <0.2 nm) and indium tin oxide (ITO) substrate (RMS surface roughness <0.7 nm) forming a vertical metal (conductive-AFM tip)-semiconductor-metal device. We have observed that the current increases as the number of layers increases up to 5 layers. By applying Fowler-Nordheim tunneling theory, we have determined the barrier heights for different layers and observed that the barrier height decreases as the number of layers increases. Using density functional theory (DFT) calculation, we successfully demonstrated that the barrier height decreases as the layer number increases. By illuminating the TMDs on a transparent ultra-flat conducting ITO substrate, we observed a reduction in current when compared to the current measured in the dark, hence demonstrating negative photoconductivity. Our study provides a fundamental understanding of the local electronic and optoelectronic behaviors of TMD-metal junction, which depends on the numbers of TMD layers, and may pave an avenue toward developing nanoscale electronic devices with tailored layer-dependent transport properties.

## INTRODUCTION

Layered transition metal dichalcogenides (TMD) are van der Waals crystal and provide the tantalizing prospect of miniaturizing semiconductor devices to truly atomic scales and accelerating the advances of many two-dimensional (2D) optoelectronic devices.[1-6] These atom thick layered TMDs demonstrate some unique properties that includes 2D confinement, direct band-gap nature[7], varying band structure with layer thickness,[7-9] and weak screening of charge carriers enhancing the light-matter interactions[7, 8, 10]. High photon absorption ($\alpha \sim 10\%$ for visible light and $\alpha \sim 40\%$ for ultraviolet photons)[7, 11], and exciton formation (a hydrogenic entity made of an *e-h* pair)[7, 11] make TMDs very attractive for different optoelectronic applications[1, 12-16] including sensitive photodetectors,[12, 17-19] energy harvesting devices,[20-22] monolayer light emitting diodes (LEDs),[13, 15, 16] single photon sources,[23-27] and nanocavity lasers.[28]

Electronic and optoelectronic devices require a metal junction for injection and collection of current. Interfacial charge transport properties play a critical role in governing the current injection or collection in a metal-semiconductor junction. One property that controls charge transport across the junction is the barrier height due to the conduction band offset between the TMD and metal. Although significant progress has been made in demonstrating many different types of TMD-based devices, it is not clear how the barrier heights change with TMD thickness at the few layers level (1L-5L).[9] One standard technique to elucidate the barrier height is through the measurement of the electron affinity by ultraviolet photoelectron spectroscopy (UPS). However, UPS requires a large sample size that is challenging to prepare by exfoliation of $MoS_2$.

One possible alternative experimental route employs surface probe microscopy (SPM) to study local electrical transport and optoelectronic properties of few layer TMDs. Since these 2D TMDs materials conform to the surface roughness of the substrate, SPM requires that the sample resides on a surface with sub-nanometer roughness (root mean square (RMS) surface roughness < 1 nm). To understand the dependence of the barrier height on TMD layer thickness, we performed conductive atomic force microscopy (CAFM) and photocurrent atomic force microscopy (PCAFM) of few-layer $MoS_2$ samples immobilized on ultra-flat transparent indium tin oxide (ITO) and template-stripped Au conducting surfaces.

To probe the layer dependence of interfacial transport properties, CAFM and PCAFM measurements of few layer $MoS_2$ samples with thickness varying from monolayer (1L-$MoS_2$) to 5 layers (5L-$MoS_2$) were performed. To understand the effect of the tip on the electrical and optical properties, Pt/Ir tips and platinum silicide (Pt/Si) tips were used for both CAFM and PCAFM measurements. Similar results were observed for both Pt/Ir tips and Pt-Silicide (Pt/Si) tips. Here we present the results obtained using Pt/Si tips unless mentioned otherwise. We have studied three $MoS_2$ samples on Au substrate and three $MoS_2$ samples on ITO coated substrate. All samples behaved similarly with respect to their electronic and optoelectronic properties. Several important features characteristic of the metal substrate/$MoS_2$/Pt-Si (tip) heterojunctions were observed; (i) the current increases as the layer number increases; (ii) the *I-V* curve analysis employed Fowler-Nordheim tunneling theory to give barrier heights that depended on the number of layers; (iii) the barrier height also depends on the type of the conducting substrate; (iv) the current was significantly lower along the edges of the $MoS_2$ basal plane; and (v) few layer $MoS_2$ samples demonstrates negative photoconductivity when illuminated by blue light (460 -490 nm). The barrier heights for 1L-5L $MoS_2$ on Au (111) surface were calculated with DFT and followed the observed experimental trends.

## RESULTS AND DISCUSSION:

Figures 1(a) and 1(b) show an optical image and the coresponding contact mode AFM height profile image of exfoliated $MoS_2$ on a freshly peeled template-stripped (TS) Au surface. Template-stripped Au substrate was prepared following standard procedures[29-32] (see Methods for details). The 1L-5L layer $MoS_2$ sheet structure is clearly observed in the contact mode AFM heigh profile image, Fig. 1(b). Before carrying out CAFM and PCAFM measurements, the RMS roughness of the TS-Au and ITO substrates were measured and found to be 0.2 nm and 0.7 nm, respectively. The ultra-smooth nature of the TS-Au surface is very close to the surface roughness of hexagonal boron nitrides (*h*-BN),[33] a layered material commonly used as an atomically flat substrate. For comparison, the RMS roughness of the external surface of the thermally evaporated (TE) gold layer was 2 nm,. The left and right insets to Fig. 1(c) show tapping mode height profile images of the TS-Au and ITO coated surfaces, respectively (see supplementaly information Fig.S1 for an AFM image of a TE coated AU surface). Fig.1(c) also compares the height histograms of the TS and TE gold surfaces confirming that template stripping produces a uniformily smooth surface with a narrow full-width-half-maximum (FWHM) distribution while the distribution and roughness of the TE surface are an order of magnitude larger. We have found that employing an ultra-flat conducting surface as a substrate for exfoliated TMD is critical to observe consistent electrical and optoelectronic properties of atomically thin MoS$_2$ crystals.

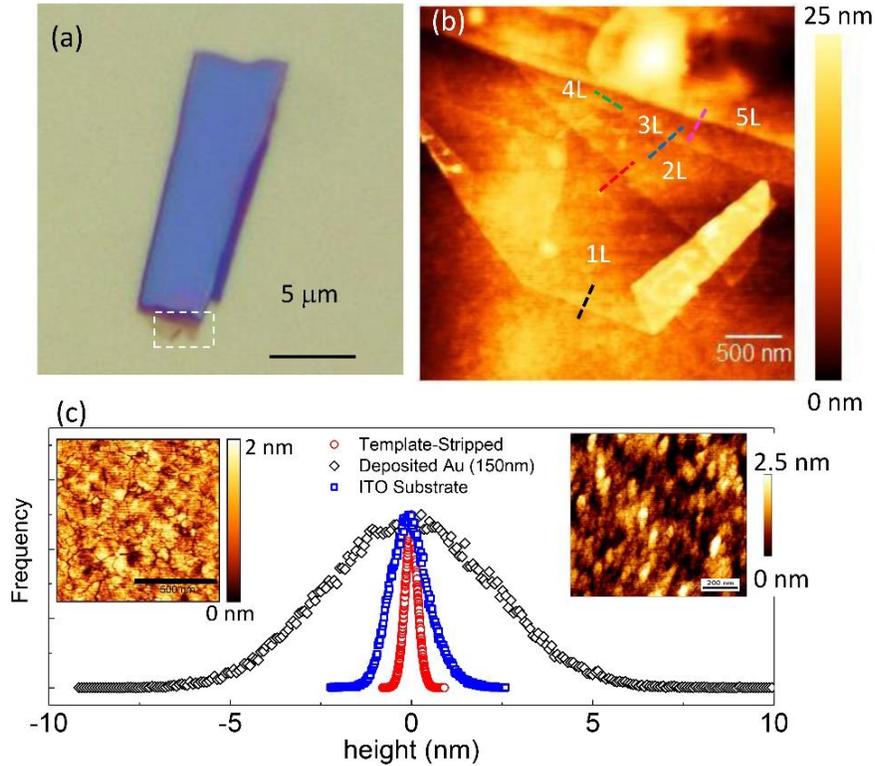

Figure 1: MoS$_2$ sample on a template-stripped Au substrate. (a) Optical image of a MoS$_2$ sample. The dotted square region is studied using AFM and CAFM. (b) The height profile AFM image of the marked area (white dotted-square) in Fig. (a) measured by AFM contact mode. Region of different layer number is shown. The scale bar is 500 nm. (c) Histogram of the height distribution (surface roughness measured by AFM) of the template-stripped Au substrate (TS-Au, ~150 nm thickness), as thermally evaporated Au (TE-Au) substrate of thickness ~150 nm and ITO substrate. The Au for the template stripping was deposited on a single crystal Si wafer. The root mean squared (RMS) value of the surface roughness measured was ~ 0.18 nm for template-stripped Au and ~1.8 nm for as deposited substrate. Inset-left: tapping mode AFM image of the template-stripped Au surface. Inset-right: tapping mode AFM image of 20 nm ITO coated substrate on a single crystal quartz substrate. The area of the scanned region is 1 μm× 1 μm. The scale bar is 500 nm for left image and 200 nm for the right image. A Gaussian fit to the height data provides full width at half maxima (FWHM) ~ 0.3 nm for the template-stripped Au, ~ 5 nm for as deposited Au and ~ 1 nm for the ITO substrate. The surface roughness profile of template-striped Au is very similar to surface roughness observed for *h*-BN (see text).

Now we discuss the electronic transport behavior of the MoS$_2$ sample on template-stripped Au substrate. Fig. 2(a) shows schematically the experimental setup used to measure the current under an applied voltage. The bottom inset shows the configuration of the sample with respect to the SiO$_2$/Si substrate. The MoS$_2$ sample was directly micro-exfoliated onto the TS-Au surface. Fig. 2(b) presents the DC current map measured with a 0.4 V bias of the sample whose height profile is shown in Fig. 1(b). The spatial current map clearly demonstrates that the current increases as the layer number increases. This feature becomes clearer in the *I-V* curves obtained by averaging over a specific layer number as shown in Fig.2(c). Almost zero current was observed for the Au substrate outside the MoS$_2$ flake (Fig.1b). We attribute the non-conducting behavior observed for the gold substrate to two factors, the small force constant (~0.2 N/m) of the cantilever and a non-conductive film, mainly hydrocarbons, that adsorbed to the gold surface within a short period of time after revealing the fresh TS surface.[34] The force constant of the tip was too small to scratch the surface contaminants to make a direct contact to the Au surface.

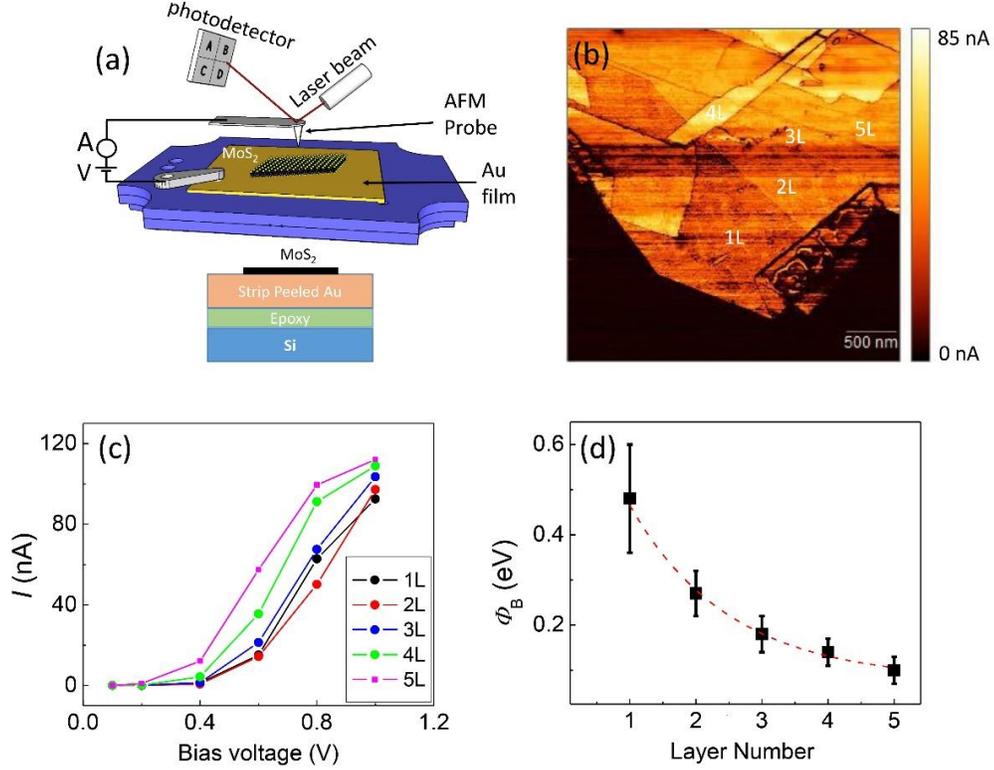

Figure 2: Layer dependent conductive AFM measurement of MoS$_2$ sample on a template-stripped Au substrate. (a) (Top) Schematic diagram of the experimental setup (not to scale) used in CAFM measurements. The current is measured by a current-amplifier. (a) (Bottom) Schematic diagram of the sample (see text for details). The MoS$_2$ sample was directly exfoliated on a template-stripped Au substrate. (b) The conductive AFM data for the sample, whose height profile is shown in Fig.1 (b). (c) Current-voltage (*I-V*) curves for different layers obtained by averaging the current for a flake. (d) The value of the barrier height measured by taking the average of every pixel. The red line is a guide to the eye. The error bar presents the standard deviation of the barrier heights.

Since the I-V curves exhibit non-linear behavior, we used Fowler-Nordheim (FN) tunneling theory to understand the layer-dependent electronic transport behavior. FN theory has been widely used to explain the tunneling of an electron between two metals separated by an insulator or semiconductor.[35-38] The tunneling current through a thin semiconductor is given by

$$I(V) = \frac{A_e q^3 m V^2}{8\pi h \Phi_B d^2 m^*} \exp\left(-\frac{8\pi\sqrt{2m^*}\Phi_B^{3/2} d}{3hqV}\right) \qquad (1)$$

where $A_e$ is the effective contact area, $h$ is Planck's constant, $q$ is the electron charge, $d$ is the thickness of the barrier, $\Phi_B$ is the barrier height, $m$ is the electron mass, and $m^*$ is the electron effective mass inside the semiconductor. We used $\frac{m^*}{m}$ ~0.35 and 0.53 for monolayer MoS$_2$ and bulk MoS$_2$ (>1L), respectively.[35] If the layer thickness $d$ is known, we can apply Eq. (1) to determine the barrier height from the *I-V* curves. Since the conducting substrate has surface roughness much smaller (~ 0.2 nm) than the thickness of monolayer MoS$_2$ (~ 0.7 nm), the separation between the bottom conducting surface and AFM tip remain constant for a specific layer number.

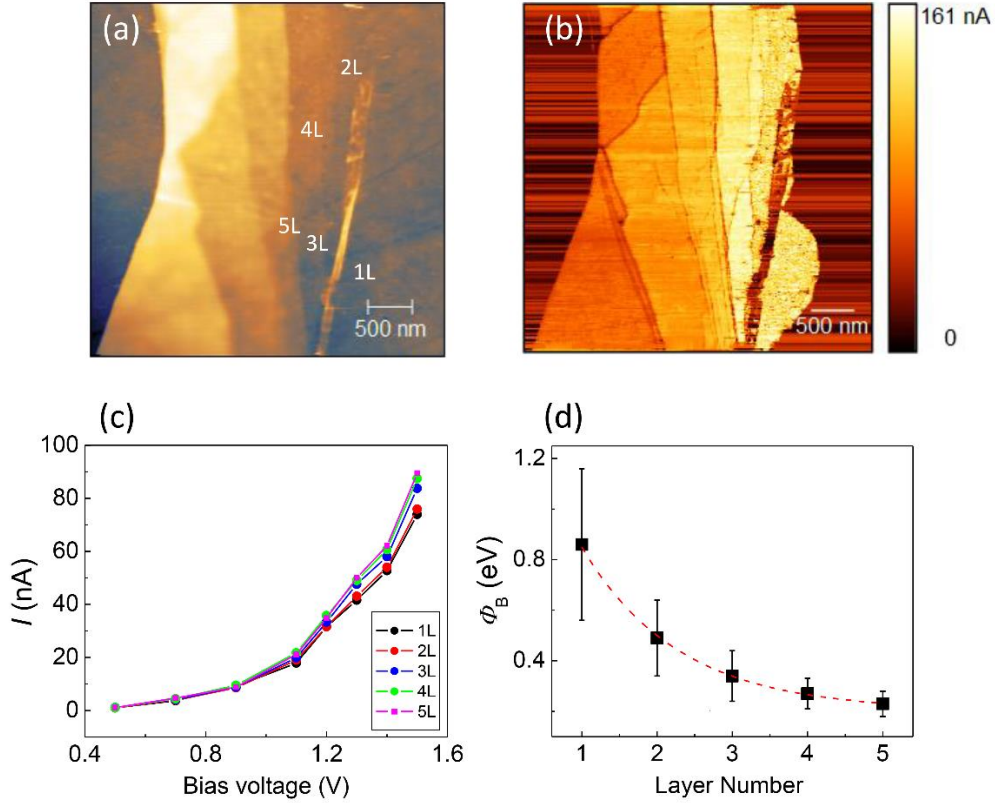

Figure 3: Layer dependent conductive AFM measurement of MoS$_2$ sample on a physical vapor deposited ITO film. (a) Contact mode AFM image of the height profile of a sample on ITO coated wafer. The layer numbers are shown. The scale bar is 500 nm. (b) The conductive AFM data for the sample for 1.5 V. (c) I-V curves for different layers obtained by averaging the current. (d) Calculated average value of the barrier height measured by taking the average of every pixel. The red line is a guide to the eye. The error bar presents the standard deviation of the barrier heights.

By using a multiple of 1L-MoS$_2$ (~0.7 nm) as the layer thickness, the barrier heights for different numbers of layers were calculated as shown in Fig.2(d). Interestingly, we observed that the barrier height decreases as the layer number increases. This suggests that the current increases as the number of layer increases, which is clearly evident both in the current map in Fig. 2(b) and the *I-V* curves in Fig.2(c). Our observations contrast with the findings of Son *et. al.*,[35] who reported that the barrier height increases as the number of MoS$_2$ layers increases for a sample immobilized on a rough ITO surface.

To understand the impact of the substrate on the barrier heights, we also studied MoS$_2$ samples on ultra-flat ITO coated single crystal quartz substrate. We found that commercially available ITO coated substrate has high surface roughness and is not suitable for SPM measurement (see supplementary Figure S1 for details). The indium tin oxide (ITO) thin films were deposited by physical vapor deposition onto transparent Z-cut quartz substrates (see Methods for details). An AFM surface profile image of the substrate is shown in the right inset to Fig.1c. The surface RMS roughness of this transparent ITO coated sample is ~ 0.7 nm.

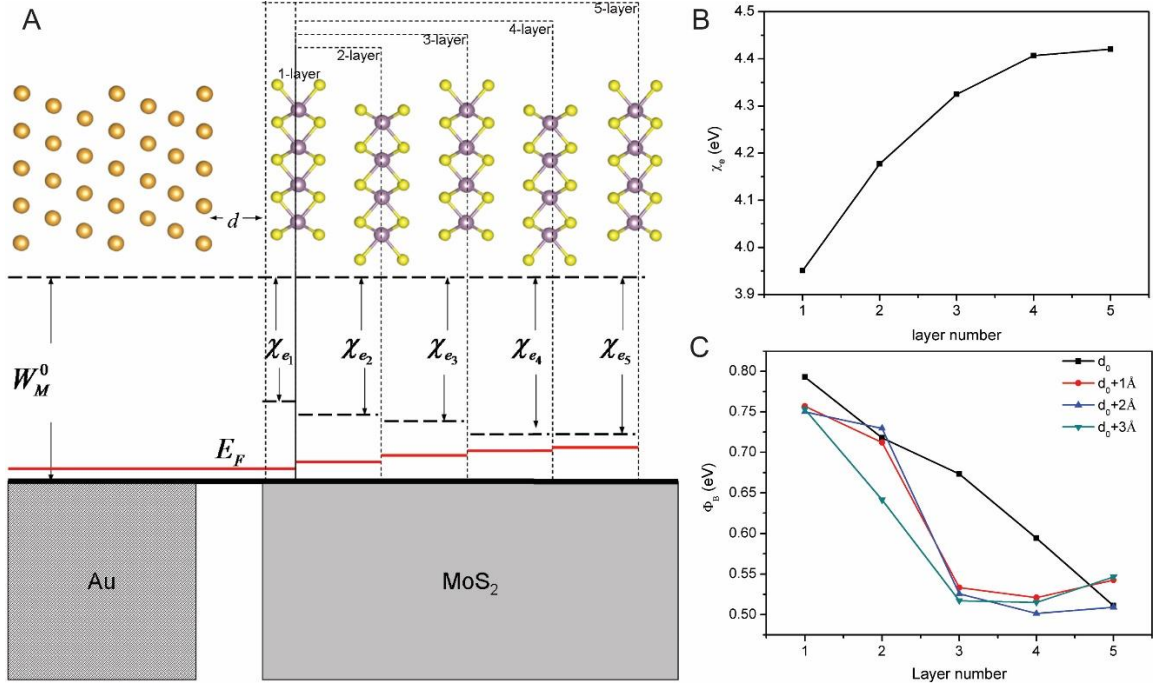

**Figure 4. DFT calculated interfacial electronic structure of the Au/MoS$_2$.** (A) Schematic of the MoS$_2$/Au interface and band alignment. (B) DFT-calculated electron affinity of MoS$_2$ with different thickness. (C) DFT-calculated Schottky barrier at the interface of Au/MoS$_2$ with varied the thickness of MoS$_2$ and interfacial spacing. The black curve shows the values from DFT-optimized structures, while the red, blue and green ones show the values when the structures are artificially displaced by 1, 2 and 3 Å further way from the Au surface, respectively.

The height profile of one MoS$_2$ sample on ITO substrate is shown in Fig. 3(a). The current map for 1L-5L MoS$_2$ at 1.5 V bias voltage is shown in Fig. 3(b) and clearly shows that the current increases with increasing layer number. The corresponding *I-V* curves are shown in Fig.3(c), which demonstrate that the current level increases, though minimally compared to MoS$_2$ on TS-Au, as the layer number increases from 1L to 5L. Note that the current level decreases above five layers. One possible explanation to account for this observation is that the barrier height does not change significantly with MoS$_2$ layer number greater than 5 layers, whereas, the sample thickness increases linearly. Hence the tunneling current decreases significantly as the sample thickness (*d*) icreases as predicted by the FN-Tunneling model. Further study is necessary to understand the details of the electrical transport properties of MoS$_2$ samples thicker than 5 layers. Fowler-Nordheim tunneling theory was used to calculate the barrier heights as shown in Fig. 3(d). Although, the layer number dependence of the barrier height for MoS$_2$ on ITO is very similar to the samples immobilized on a TS-Au substrate, the barrier height on the ITO substrate is much higher than the barrier heights of MoS$_2$ on the TS-Au substrate, which is likely caused by different work function of ITO and Au in the presence of multilayer MoS$_2$.

Another important feature observed in these MoS$_2$ sample (both on TS-Au and ITO substrate) is that the edges are less conducting compared to the middle of the basal plane. The current drops by 2-5 times as the tip approaches the edge compared to regions far from the edges. The low current edges are especially visible, both in Fig.2(b) and Fig. 3(b), as dark lines that trace the boundaries between adjacent layers. The current profile along three different edges are presented in the supplement Fig. S2. Our estimate of the width of that insulating edge is ~20 nm, which is comparable to tip Pt/Si tip width. This suggests that the insulating region is smaller than Pt/Si tip and is beyond our measurement capabilities with our current AFM probe and setup.

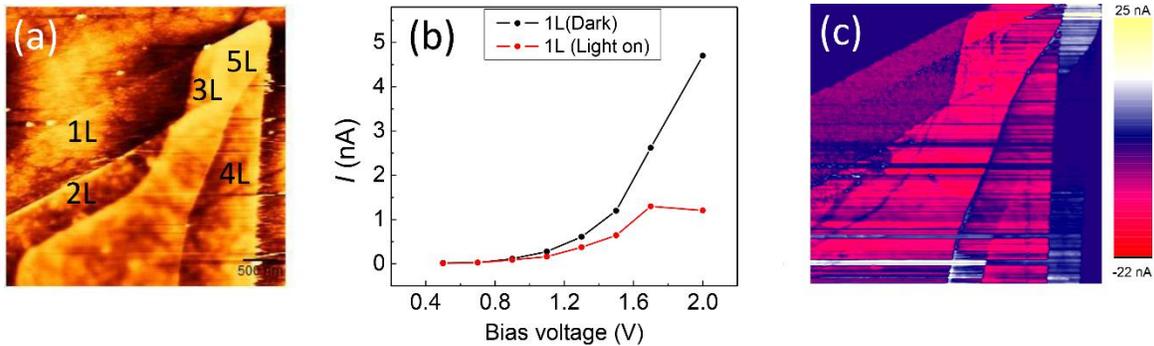

Figure 5: Photoconductive AFM measurement of sample on an ultra-flat ITO coated single crystal quartz. (a) Contact mode AFM image of the height profile of the sample. The monolayer region is marked. (b) The IV curve when the sample is illuminated by blue light (black circle) and in the dark (red circle). (c) The photoconductive AFM image of the sample at V=1 V. The image is obtained by subtracting the illuminated photoconductive AFM image from the conductive AFM image measured in the dark.

To explain our experimental results, density functional theory (DFT) calculations were performed to explore the interfacial electronic structure of 1L-5L $MoS_2$ on a Au(111) surface (Fig.4A). Figure 4B depicts the calculated electron affinity ($\chi_e$) of isolated $MoS_2$ layers, which shows an incremental increase of the electron affinity (lower conduction band minimum) with increased layer number. The electron affinity reaches a plateau at about 4 layers. Figure 4C shows the calculated Schottky barrier, which has been calculated as $\phi_B = W - \chi_e$, where $W$ is the work function of gold. The calculated work function is shown in Figure S3 and shows a systematic decrease with layer thickness for the optimized structures. Fig. 4c also shows the effect of increasing the interfacial distance between the $MoS_2$ layers and substrate. The imposed interfacial separation accounts for the possibility that the experimental interfacial spacing may not be exactly the same as the DFT-optimized ones (black curve in Fig.4C) due to the corrugation of the Au surface as well as contaminants at the interface. The calculated Schottky barriers for an interlayer spacing of 1-3 Å between Au and $MoS_2$ shows the general trend of decreasing barrier with increasing layer number, although $\phi_B$ reaches a plateau after 3 layers, which is more in line with the experimental results.

Finally, we discuss the optoelectronic behavior of exfoliated 1L-5L $MoS_2$ sample on an ultra-flat ITO substrate. The $MoS_2$ (1L-5L) samples were immobilized on a transparent ITO substrate. The AFM stage was mounted on an inverted Zeiss microscope. The $MoS_2$ was illuminated through the quartz substrate with an LED source (X-Cite® 110), which employed a dichroic filter to select the excitation wavelength range of 460-490 nm. The excitation light was guided by a microscope objective (x20, NA~0.75) to the sample. The beam area was ~10 μm with optical power density $P \sim 4$ mW/μm². The height profile of the sample is shown in Fig. 5(a) where different layer regions are marked. The corresponding $I$-$V$ curve for 1L-$MoS_2$ is shown in Fig. 5(b). Surprisingly, the current decreases as the sample is illuminated by 460-490 nm photons demonstrating negative photoconductivity. Note that the change in photocurrent ($\Delta I_{PC}$) become larger as we increase the bias voltage. The difference in current under illumination $\Delta I_{PC} (= I_{light} - I_{dark})$ for $V_{ds} = 1$ V is shown as a spatial map in Fig. 5(c). The spatial map was made by subtracting the AFM current image acquired without illumination from the one illuminated with blue light and clearly shows that the current decreased under illumination. Another interesting feature is that $\Delta I_{PC}$ becomes more negative as the layer number increases. The observation of negative photoconductivity of single crystal $MoS_2$ nano-sheet is an anomalous photoresponse compared to regular positive photoconductivity behavior observed for planar structured TMDs.[2, 4, 11, 12]

Negative photoconductivity of MoS$_2$ nano-sheet connected by two metal electrodes in a planar or horizontal configuration was observed previously and the effect was attributed to the strong many-body interactions in MoS$_2$.[39, 40] This many-body related negative photoconductivity in planer MoS$_2$ device configuration has transient nature with lifetimes on the order of trions (~picoseconds). On the other hand, negative photoconductivity observed in our vertical metal/semiconductor/metal device structure is non-transient or time-independent, which suggests that the origin of negative photoconductivity in our devices is completely different than planar device configurations. One possible explanation is that the excited electron and holes after optical illumination in MoS$_2$ modify the barrier height, which therefore changes the tunneling current. Strong negative photoconductivity measured across the basal plane suggests that the barrier height increases due to optical illumination, which in turn reduces the tunneling current. Because photoconductivity is a critical parameter applied to optoelectronics, further study is necessary to elucidate the origin of this DC negative photoconductivity.

## CONCLUSIONS:

In conclusion, electrical and optoelectronic properties of 1L-5L MoS$_2$ samples residing on two different atomically flat conducting surfaces are probed by using CAFM and PCAFM measurements. We observed four important features of electrical and optoelectronic properties of MoS$_2$ nanosheets measured perpendicular to the basal plane. First, Fowler-Nordheim tunneling theory shows that the barrier heights correlates with the number of layers; the barrier height is highest for the monolayer and then monotonically decreases as the layer number increases. By using DFT calculations, we attributed this observation to the decrease of the electron affinity as the layer number increases. Second, the 1L-5L MoS$_2$ barrier height depends on the type of the conducting substrate underneath the sample; gold has a lower barrier height than ITO. Third, the edges of MoS$_2$ are less conducting than the basal plane. And fourth, negative DC photoconductivity was observed by illuminating the sample with blue light ($\lambda \sim$ 460-490 nm). Our study revealed the interfacial electrical and optoelectronic properties while in contact with an ultra-flat conductor and can lead to the development of nanoscale optoelectronic devices with tailored properties.

## ACKNOWLEDGEMENT:

The authors acknowledge the support from San Francisco State University (SFSU), the Center for Computing for Life Sciences at SFSU and to the NSF for instrumentation facilities (NSF MRI-CMMI 1626611). H.L, V.Z.C. and A.K.M.N. acknowledge the support from the National Science Foundation Grant ECCS-1708907 and Department of Defense Award (ID: 72495RTREP). The computational research used the supercomputer resources of the National Energy Research Scientific Computing Center (NERSC) and the OU Supercomputing Center for Education & Research (OSCER) at the University of Oklahoma. S.D. and E.P. acknowledges support from the Stanford Non-volatile Memory Technology Research Initiative (NMTRI) and partial support from SRC grant 2532.001.

## METHODS:

***Template-stripped gold substrate:*** Prime grade silicon (100) wafers were cleaned with a solution of water, ammonium hydroxide, and 30 wt% hydrogen peroxide in a ratio of 4:1:1 at a temperature of ~70 °C for ten minutes (Standard Clean-1). The wafers were then rinsed with copious deionized water, dried under argon, and then placed in a bell jar. The bell jar was evacuated to a base pressure of 10$^{-7}$ torr and then 150 nm of gold was thermally evaporated onto the silicon at a rate of 5 Å/s. After the wafers were removed from the bell jar, precut silicon pieces with dimensions 1 cm x 1 cm were adhered to the gold surface with Devcon 2-Ton® Epoxy. To dry the epoxy and make a strong attachment, the diced wafer/epoxy/deposited-Au-film were kept under pressure for 24 hours. The pressure was created by a block of heavy metal (5~7 pounds). Finally, the diced wafer was peeled/stripped using a razor blade. To avoid organic contamination of the freshly peeled gold surface, micro-exfoliated MoS$_2$ flakes from a bulk sample was transferred by direct contact with the gold surface within ~ 5 minutes after template stripping.

***Indium tin oxide (ITO) thin films preparation:*** ITO films were deposited with physical vapor deposition onto transparent Z-cut quartz substrates. The films were sputtered from an as-purchased ITO target in a 10

mTorr Ar ambient at 65 W DC plasma deposition power at room temperature, with a deposition rate ≈ 3 nm/min. In this work, we have employed 20 nm thick ITO coated film. The samples were thermally annealed at 400 C to reduce the sheet resistance (~220 Ω square).

***Conducting AFM (CAFM) and photoconductive AFM (PC-AFM) measurements:*** CAFM and PC-AFM measurement were conducted using a JPK Nanowizard® 4a Bioscience integrated on an inverted Zeiss Fluorescence microscope. JPK conductive AFM module was used to conduct CAFM and PC-AFM measurements. The total AFM system was mounted inside a JPK acoustic enclosure, which was also optically opaque, Hence the electrical measurement was conducted in a totally dark environment. For PC-AFM measurement, the light was guided by a liquid light guide from an LED light source (X-Cite® 110 LED) into a microscope objective (x20, NA~0.75). The blue illuminating light was filtered by a Zeiss dichroic filter (Excitation BP 470/40). AFM images were analyzed and plotted using both Gwyddion software package[41] and JPK data processing software package. The Pt-Si (PtSi-CONT, resonant frequency, $f$~13 kHz, force constants $k$~0.2 N/m) and Pt/Ir (ANSCM-PC, f~ 12 KHz and k~0.2 N/m ) probes tips were purchased from Nanoandmore USA and Applied NanoStructures, Inc., respectively.

***Density functional theory (DFT) calculations***: The DFT calculations were performed within the framework of plane-wave method as implemented in the VASP code.[42] The exchange and correlation potential was described by using the generalized gradient approximation (GGA) in the scheme of Perdew-Burke-Ernzerhof (PBE) functional[43] and the projector-augmented wave method[44,45] was employed to give the numerical description of the ion-electron interaction. A cutoff energy of 400 eV was used to limit the plane-wave basis set. The equilibrium distances between the optimized Au substrate and $MoS_2$ with different layer numbers were determined by calculating the minimums of the energy versus distance curves as relaxed under the stop criterion of 0.03 eV/Å per atom. A vacuum layer of 40 Å has been added to each composite structure of Au and MoS2 to simulate the surface configuration. The Van der Waals (vdW) correction based on the Grimme's DFT-D2 method[46] and dipole correction along the direction vertical to the interface have been included to calculate the total energies and electrostatic potentials, which can provide the reasonable atomic configurations at the interfaces and the vacuum potentials.